\documentclass[%
reprint,
superscriptaddress,
 amsmath,amssymb,
 aps,
floatfix,
]{revtex4-1}

\usepackage{graphicx}
\usepackage{float}
\usepackage{braket}
\usepackage{subcaption}
\usepackage{color, ulem}
\usepackage{gensymb}
\usepackage{bm}

\begin{document}

\title{Moir\'e Modulation of Charge Density Waves }

\author{Zachary A. H. Goodwin}
\email{zachary.goodwin@manchester.ac.uk}
\affiliation{National Graphene Institute, University of Manchester, Booth St. E. Manchester M13 9PL, United Kingdom\\}
\affiliation{School of Physics and Astronomy, University of Manchester, Oxford Road, Manchester M13 9PL, United Kingdom\\}

\author{Vladimir I. Fal'ko}
\email{vladimir.falko@manchester.ac.uk}
\affiliation{National Graphene Institute, University of Manchester, Booth St. E. Manchester M13 9PL, United Kingdom\\}
\affiliation{School of Physics and Astronomy, University of Manchester, Oxford Road, Manchester M13 9PL, United Kingdom\\}
\affiliation{Henry Royce Institute for Advanced Materials, University of Manchester, Manchester M13 9PL, United Kingdom\\}

\date{\today}

\begin{abstract}
Here we investigate how charge density waves (CDW), inherent to a monolayer, are effected by creating twisted van der Waals structures. Homobilayers of metallic transition metal dichalcogenides (TMDs), at small twist angles where there is significant atomic reconstruction, are utilised as an example to investigate the interplay between the moir\'e domain structure and CDWs of different periods. For $\sqrt{3}\times\sqrt{3}$ CDWs, there is no geometric constraint to prevent the CDWs from propagating throughout the moir\'e structure. Whereas for $2\times2$ CDWs, to ensure the CDWs in each layer have the most favourable interactions in the domains, the CDW phase must be destroyed in the connecting domain walls. For $3\times3$ CDWs with twist angles close to 180$\degree$, moir\'e-scale triangular structures can form; while close to 0$\degree$, moir\'e-scale dimer domains occur. The star-of-David CDW ($\sqrt{13}\times\sqrt{13}$) is found to host CDWs in the domains only, since there is one low energy stacking configuration, similar to $2\times2$ CDWs. These predictions are offered for experimental verification in twisted bilayer metallic TMDs which host CDWs, and we hope this will stimulate further research on the interplay between the moir\'e supperlattice and CDW phases intrinsic to the comprising 2D materials.
\end{abstract}

\maketitle

\section{Introduction}

Twistronics~\cite{Carr2017twist,Carr2020Rev} is the study of how the relative twist angle in van der Waals heterostructures of 2D materials effects the systems properties~\cite{moiresim}. This field gained significant attention upon the discovered of correlated insulating states and superconductivity in the so-called magic-angle ($\sim1.1\degree$)~\cite{Bistritzer2010} twisted bilayer graphene (tBLG)~\cite{NAT_I,NAT_S}. Since then, magic-angle tBLG has been reported to host strange metallic behaviour~\cite{SMTBLG,Polshyn2019}, nematic order~\cite{NAT_MEI,NAT_CO,IEC,Cao2020}, Chern insulating states~\cite{Chern_Wu,Stepanov2021zero,Xie2021frac,Chern_Nuckolls,Pierce2021}, Dirac revivals~\cite{Zondiner2020,Wong2020}, Pomeranchuk effect~\cite{Rozen2021,Saito2021}, amongst other phases and effects, as summarised in Refs.~\citenum{Balents2020rev,Andrei2020Rev}. In addition, the tBLG devices are highly tunable, through the twist angle~\cite{NAT_I,NAT_S}, doping level from electrostatic gating~\cite{SOM}, applied hydrostatic pressure~\cite{TSTBLG,PDTBLG}, and through aligning with the substrate~\cite{EFM,Serlin2020,Tschirhart2021} or from changing its thickness~\cite{PHD_3,Stepanov2020,Saito2020,Vafek2020}.

Shortly after the initial discoveries of magic-angle tBLG, twistronics progressed towards creating moir\'e structures out of 2D materials other than graphene, such as twisted homobilayers of semiconducting transition metal dichalcogenides (TMDs)~\cite{Carr2017twist,Naik2018,Zhang2020TMD,Vitale2021}. In twisted bilayers of WSe$_2$, correlated insulating states were found at angles of $4-5\degree$ when one electron is removed per moir\'e unit cell~\cite{wang2020correlated,Ghiotto2021}, in addition to some signatures of superconductivity~\cite{wang2020correlated}. Twisted semiconducting TMDs are also interesting because of their excitonic states which can be tuned through the relative twist angle~\cite{Andersen2020exc,Merkl2020}, and their ferroelectric properties which allow control over the domain sizes with perpendicular electric fields~\cite{Weston2022,Enaldiev2022Ferro}.

Creating moir\'e materials out of weakly correlated 2D materials, such as graphene and semiconducting TMDs, induces emergent correlated phases through the onset of flat bands~\cite{moiresim}. There are, however, monolayers which are \textit{already} strongly correlated and intrinsically host broken symmetry phases~\cite{moiresim}. For example, CrI$_3$ exhibits ferromagnetic order in the monolayer limit~\cite{Huang2017}, and recently moir\'e structures of CrI$_3$~\cite{Hejazi2020,Wang2020CRI} were created~\cite{Song2021mag,Xie2022CrI3}. In $\sim0.2\degree$ twist-angle bilayers of CrI$_3$, the magnetic order was found to vary from ferromagnetism to antiferromagnetism throughout the moir\'e superlattice~\cite{Song2021mag}. Whereas, in $\sim0.9\degree$ twist-angle double-bilayers, the magnetic states were found to be linear combinations of the ordering tendencies of bilayers and quadlayers~\cite{Xie2022CrI3}. At the time of writing, this appears to be one of the only examples which utilises a strongly correlated 2D monolayer/bilayer, that intrinsically hosts a broken symmetry phase, in a moir\'e material~\cite{moiresim}.

\begin{figure*}
\centering
\includegraphics[width=0.9\textwidth]{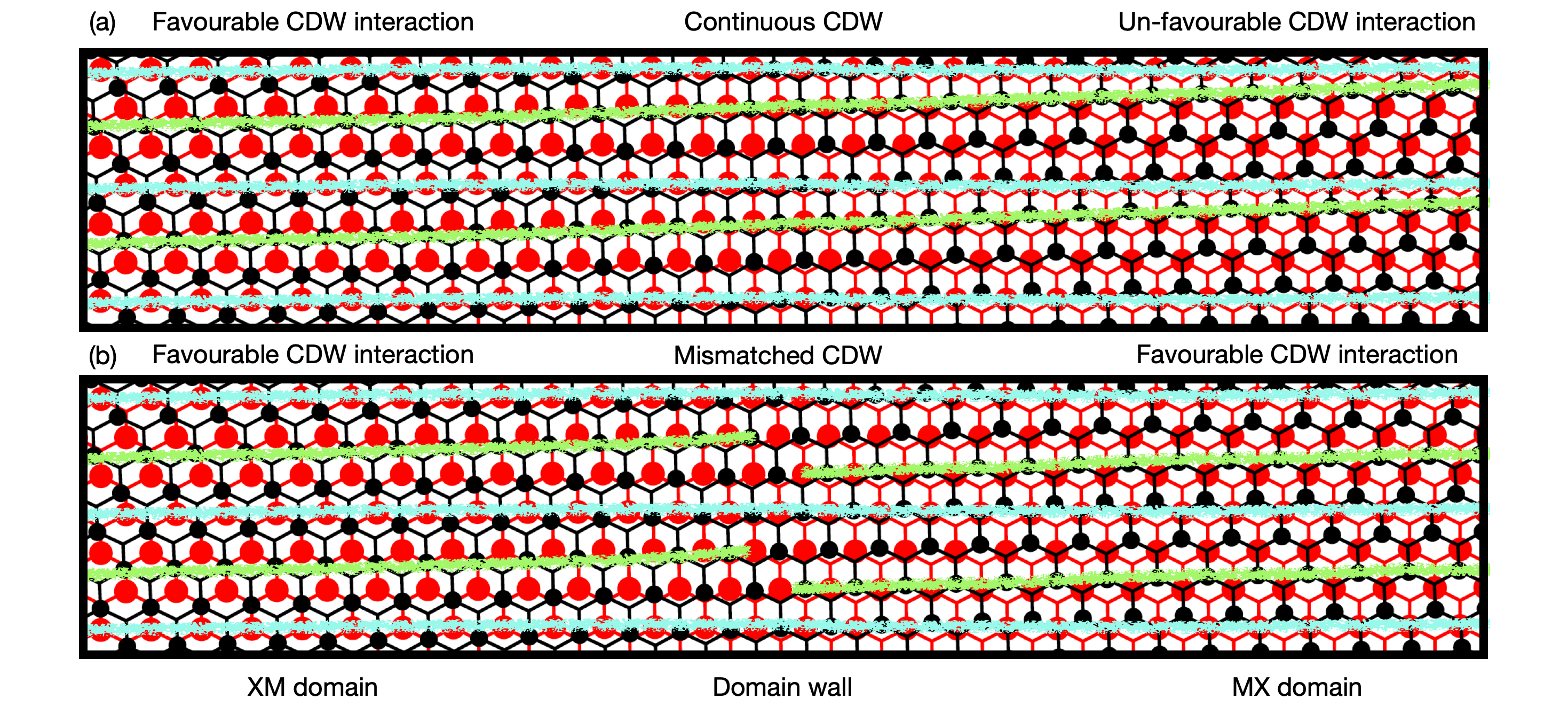}
\caption{Schematic of a 1D charge density wave (CDW) propagating between two domains in a parallel twist angle moir\'e. The red lattice is the bottom layer, and the black lattice is the top layer, with transition metal atoms denoted by filled circles (see Moir\'e Structure for more details). The maximum in the charge density of the 1D CDW in the bottom layer is denoted by a blue line, and for the top layer by a green line. On the left hand side is the XM domain, on the right is the MX domain, and between these domains is a domain wall. (a) - Corresponds to the case when the interactions between the CDWs in each layer is small and the energy cost to destroy the CDW is large. The CDWs presumably survive everywhere in each layer for this case. This causes the CDW in each layer to interact favourably in some domains and un-favourably in other domains. (b) - Corresponds to the case when there is a strong interaction between the CDW of each layer. This will cause the domains to prefer to have a favourable stacking to lower the energy. Propagating the CDWs from the domains to the domain wall demonstrates that there is a mismatch between the CDWs in the domain wall, which presumably causes the destruction of the CDW in the domain wall.}
\label{fig:single}
\end{figure*}

\begin{figure*}
\centering
\includegraphics[width=0.7\textwidth]{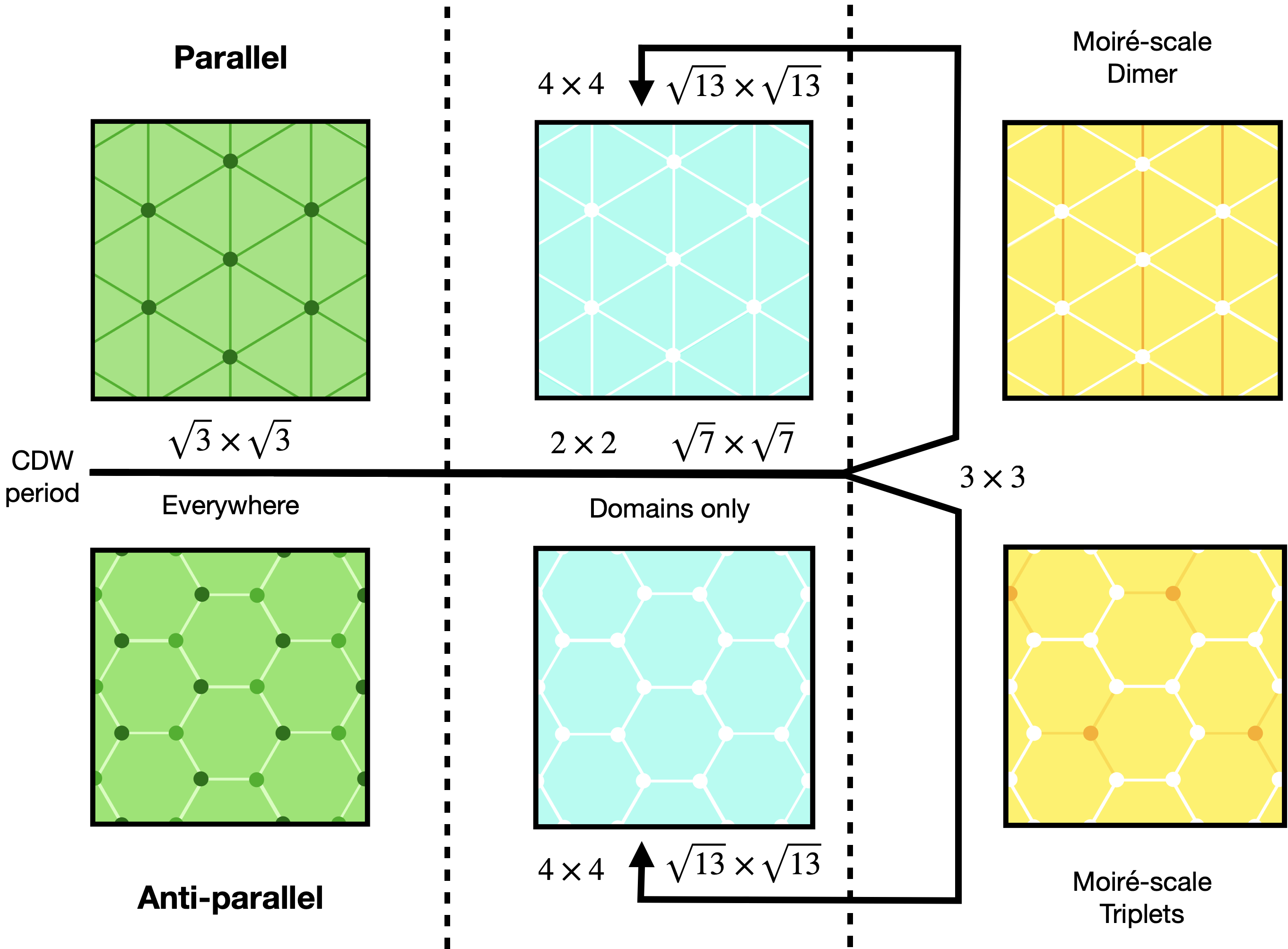}
\caption{Schematic summary of how charge density waves (CDWs) of different periods behave in moir\'e supperlattices of parallel ($\sim0\degree$) and anti-parallel twisting ($\sim180\degree$). The period of the CDW is denoted by $m\times m$, where $m$ is the length of the CDW period in units of the lattice constant of the monolayer. Parallel (P) twisting has a honeycomb lattice of triangular domains and anti-parallel (AP) has a triangular lattice of hexagonal domains; domain walls are shown by lines between these domains, and vertex regions are shown by filled circles where these domain walls connect. Further details of the domain structures can be found in the Moir\'e Structure Section. If the CDW phase survives, we denote this with a colour, and if it is destroyed, we denote this by white lines/circles. We find that the qualitative behaviour of how CDWs survive in moir\'e superlattices sensitively depends on the CDW period, but it is not very sensitive to P vs. AP twisting. Overall, we find 3 types of behaviour, as separated by dashed vertical lines: (left) - CDWs persisting everywhere, because there is no geometric or energetic constraint which should destroy it (middle) - CDWs surviving in the domains only, owing to a single low-energy stacking configuration (right) - Moir\'e-scale structures of CDWs. For P twisting, pairs of domains with the lowest energy stacking are joined, and this behaviour is referred to as moir\'e-scale dimer domains; whereas, for AP twisting, three domains with the lowest energy stacking are joined at a vertex, which are referred to as moir\'e-scale triplet domains. To observe moir\'e-scale structures (not just the CDWs existing everywhere or in domains only), there is an optimal period of the CDW, which is large enough to have several degenerate stacking configurations, but short enough that these degenerate stacking configurations are related to each other by the Burgers vector of the domain wall.}
\label{fig:summ}
\end{figure*}

%
%
%
%
%
%
%
%
%
%
%
%
%
%
%

Another class of 2D material which can host broken symmetry phases in the monolayer limit is metallic TMDs~\cite{Xu2021REV,Rossnagel2011REV}, where charge density waves (CDW) have been found~\cite{Chen2020CDW,Chen2017,Ma2016,Ryu2018,Galvis2013,Tsen2015,Chen2018,Ugeda2016,Wang2019,Chen2015,Lin2020,Nakata2016,Efferen2021}. If a twisted homobilayer of such TMDs is constructed, what would happen to the CDW phase in each monolayer? It could either be destroyed in favour of emergent moir\'e effects (not discussed here), or the moir\'e supperlattice could modulate this broken symmetry phase (either existing everywhere with its magnitude varying, or being destroyed in certain regions and existing in others). For the latter case, since the period of the CDWs is longer than the primitive lattice vector, and neighbouring domains in moir\'e supperlattices are related to each other through a Burgers vector which is of the order of the primitive lattice vector (or shorter), this suggests how the CDWs stack on top of each other can change in neighbouring domains. This raises the question: does the moir\'e supperlattice impose any constraints on where the CDWs could exist, provided the CDWs in each layer interact strongly?

Here we investigate if the moir\'e supperlattice can impose any constraints on where CDWs can exist, and therefore, how they are modulated in the moir\'e supperlattice. We focus on marginal twist angles of metallic TMD homobilayers with significant atomic reconstruction of the moir\'e superlattice, where large domains of the most favourable stacking emerges. An effective model is constructed to describe this system, based on a number of simplifying assumptions. For different periods of the CDWs phases, we investigate how the CDWs can propagate between domains. It is found that the period of the CDW can significantly alter how the CDW phase is modulated in the moir\'e superlattice.

The paper is structured as follows. First, the results of the analysis are summarised in a preview. The rest of the paper provides the details about how the results were obtained. First, the employed methodology is outlined, starting with the assumptions, which is followed by a summary of the moir\'e structures and utilised free energy. Next, the analysis for how different CDW periods are modulated in moir\'e supperlattices is presented. Finally, we discuss the limitations of the presented analysis and highlight areas of interest for future research.

\section{Preview}

For 1H monolayers, which are susceptible to CDWs, stacking and twisting these monolayers in parallel (P) and anti-parallel (AP) configurations creates moir\'e structures with distinct domain structures~\cite{Weston2020TMD,Enaldiev2020,Maity2021rec}. For P twisting, a hexagonal lattice of triangular domains connected by a partial screw dislocation occurs; while for AP twisting, a triangular lattice of hexagonal domains connected by a full dislocation occurs (see Moir\'e Structure Section for more details). Our aim is to understand how CDWs of different periods propagate throughout these moir\'e supperlattices, and see how they will be modulated assuming the CDWs in each layer interact strongly (amongst other approximations - see Methods Section for more details).

\begin{table}
\centering
\begin{tabular}{lcccc}
\hline
CDW     & $\quad$ & Anti-parallel & $\quad$ & Parallel\\
\hline
\hline
$\sqrt{3}\times\sqrt{3}$   &  & everywhere & & everywhere\\
$2\times2$    & & domains only & & domains only\\
$\sqrt{7}\times\sqrt{7}$    & & domains only & & domains only\\  
$3\times3$    & & triplet domains & & dimer domains\\
$\sqrt{13}\times\sqrt{13}$    & & domains only & & domains only\\ 
$4\times4$    & & domains only & & domains only\\ 
\hline
\end{tabular}
\caption{Summary of the results shown in Fig.~\ref{fig:summ}.}
\label{tab:summ}
\end{table}

To provide some intuition of the effective model that our results are based on, we describe how a 1D CDW propagates over a domain wall of a P moir\'e structure. In Fig.~\ref{fig:single} a 1D CDW that is parallel to the direction of the Burgers vector between each domain is shown, where the lines denote the maximum charge density of the CDW in each layer. Let us focus on the left side of Fig.~\ref{fig:single}(a) to start with, where there is a favourable interaction between the CDW of each layer because the maxima in the CDWs are out-of-phase. Going over a domain wall to a neighbouring domain causes the lattices of the monolayers to shift. Assuming the CDWs in each layer are strongly bound to the underlying lattice, i.e. commensurate CDWs, the CDWs would now interact un-favourably (the variation of the interaction energy can be modelled with a sine-Gordon equation following Ref.~\citenum{Nguyen2017} and the interaction energy given later, as done for magnetic order in Ref.~\citenum{Wang2020CRI}). This case, shown in Fig.~\ref{fig:single}(a), corresponds to CDW phases in each layer that do not strongly interact with each other, and where there is a high energy cost to destroy the CDW phase within each layer. Therefore, the CDW phase exists everywhere in each monolayer.  On the other hand, if the CDWs in each layer interact strongly with each other and there is not a large energy cost to destroy them within a layer, the CDWs will prefer to retain the energetically favourable interaction in the domains. Propagating the CDW phases towards the domain wall reveals that there is a mismatch between these CDWs, as shown in Fig.~\ref{fig:single}(b), which should destroy the CDW phase in the domain walls. This latter case of strongly interacting CDWs is the basic intuition for the effective model which is developed (see Methods Section for details of the model).

In Fig.~\ref{fig:summ} and Tab.~\ref{tab:summ}, the results of the presented analysis are summarised for different CDW periods. A difference between P and AP is only found for $3\times3$, with P and AP behaving analogously for all other studied periods. When the CDW period is $\sqrt{3}\times\sqrt{3}$, it is found that the CDW phase can exist everywhere in the moir\'e structure. This is because there are 3 degenerate stacking configurations of the CDWs, which are converted between each other upon crossing domain walls. In contrast, $2\times2$ and $\sqrt{7}\times\sqrt{7}$ exist in the domains only because there is only 1 low energy stacking configuration, and crossing domain walls shifts the stacking away from this favourable interaction. Therefore, the CDW phase must be destroyed in the domain walls, and presumably the vertex regions too. For $3\times3$, it is possible for the CDWs to propagate between certain domains without being destroyed in the domain walls, since degenerate stacking configurations are connected by the Burgers vector of the domain wall which is crossed. For AP twisting, moir\'e-scale triplet domains occur, whereas for P twisting moir\'e-scale dimer domains occur. Finally, for larger CDW periods, such as the star-of-David CDW ($\sqrt{13}\times\sqrt{13}$) and $4\times4$ CDW, it is again found that the CDW can exist in the domains only. 

Overall, it is found for very short periods, the CDWs persist everywhere; for slightly larger CDW periods, the CDWs only exist in the domains; for intermediate CDW periods, moir\'e-scale structures can occur; and large CDW periods, the CDWs only exist in the domains. Therefore, there is some optimal period for the interplay between CDWs and the moir\'e supperlattice. This occurs because there must be a number of degenerate stacking configurations which are connected to each other by the Burgers vector which describes how the layers stack in neighbouring domains. 

\section{Methods}

In the presented work, marginal twist angle homobilayer (metallic) TMDs with significant lattice relaxation are studied as a test case. The structural relaxations of metallic TMDs, which are typically 1T monolayers, in moir\'e structures is not well known. Therefore, it is assumed that the structural relaxations which occur are identical to those of the semiconducting twisted TMDs~\cite{Enaldiev2020} with 1H monolayers. In the context of twisted homobilayers of semidconducting TMDs, Enaldiev \textit{et al.} showed that the domain structures form when $\theta < 2\degree$ for P twisting and when $\theta < 1\degree$ for AP twisting~\cite{Enaldiev2020}. Therefore, the twist angle must at least be smaller than these limits, and the employed approximations should work better the smaller the twist angle, i.e. $\theta < 0.5\degree$. For twist angles larger than this, there is not significant lattice relaxations, and the CDWs in each layer won't be able to coherently interact with each other over large areas.

%
%

For this system, the following assumptions are made:


\begin{enumerate}
\item The monolayer is susceptible to the formation of a commensurate CDW. Incommensurate CDWs will be controlled by the electronic properties, and will be not modulated by the moir\'e superlattice, since they are not locked to the lattice. Furthermore, it is assumed that there is one dominant CDW period, so competing orders do not need to be considered.
    
\item The presence of the other layer is assumed not to intrinsically change the driving force to form the CDW in each layer. Therefore, we assume the CDWs in each layer are only altered through the interaction of the two CDWs, i.e. an increase in the order parameter from a favourable interaction stabilising the phases. This assumes that there is no emergent moir\'e CDW which forms from intrinsic changes to the electronic structure (formation of flat bands) or electron-phonon coupling. 

\item Metallic TMDs typically have 3 CDWs which occur together, owing to the 3-fold symmetry of the lattice. It is assumed that maxima of the 3 CDWs coincide at a point on the lattice (this is when there is zero intralayer phase shift between the components of the CDW phase). The transition metal atom is taken as the point at which the CDWs coincide. Note that using the chalcogen atoms instead does not change the predictions.

\item It is assumed that the periodic lattice distortion of the CDWs is small, so we can work with the relaxed bilayer structures. Therefore, the structural relaxations occur in the normal state, and are frozen. Then how the CDWs survive on this moir\'e superlattice is investigated.

\item The CDWs in each layer strongly interact, which causes them to prefer stacking configurations which are out-of-phase. We assume in each domain the most favourable interaction between the CDWs in each layer always occurs. 

\item It costs energy to destroy the CDW, and it is assumed it can only be destroyed once when transitioning from each domain, i.e. it cannot be destroyed between domains and domain walls to keep the CDW phase in the domain walls. 
   
\item The CDW phase in each monolayer has a coherence length longer than the moir\'e length scale. It is also assumed that there is no external strain, i.e. from a varying twist angle, since strain can alter CDWs.
    
\end{enumerate}

\begin{figure*}
\centering
\includegraphics[width=0.9\textwidth]{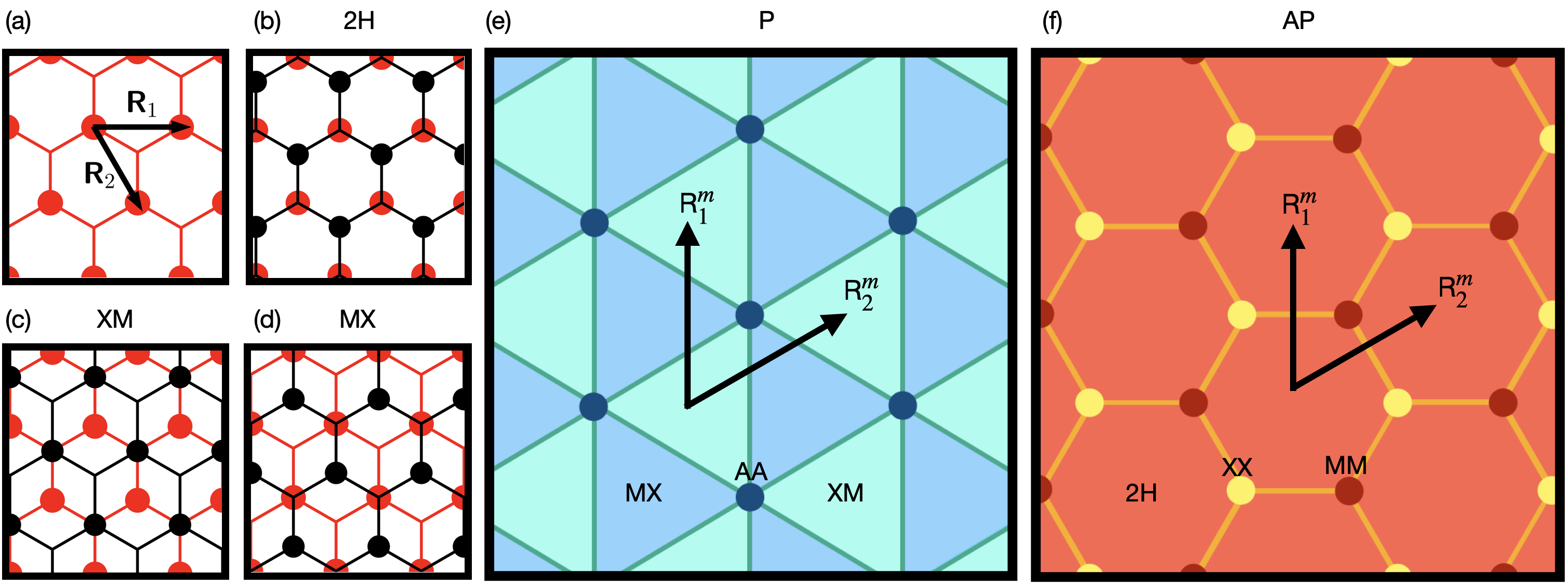}
\caption{Schematics of the utilised structures. (a) - schematic of 1H monolayer TMD, where the transition metal atom is denoted by the filled red circle and the chalcogens are located where the three lines cross. Lattice vectors utilised are shown. (b)-(d) schematics of bilayer stacking which occur in the domains. Top layer is shown in black and bottom layer in red. (e) schematic of honeycomb lattice of triangular domains of marginal twist angles close to 0$\degree$, referred to as the parallel (P) twisting. The MX and XM domains are indicated, as well as the AA stacking regions of the vertex, and domain walls separate the XM/MX domains and join at the AA regions. The moir\'e lattice vectors are indicated. (f) - schematic of triangular lattice of hexagonal domains of 2H stacking of marginal twist angles close to 180$\degree$. Domain walls shown in a different colour, and the two different vertex regions are shown.}
\label{fig:Struct}
\end{figure*}

\subsection{Moir\'e structure}

The 1H monolayer polytype is assumed. Several 1H structures exhibit CDWs, such as NbSe$_2$~\cite{Ugeda2016} and TaSe$_2$~\cite{Ryu2018} which both form $3\times3$ CDWs, but there are many cases in the 1T structure~\cite{Chen2020CDW,Chen2017,Ma2016,Galvis2013,Tsen2015,Chen2018,Wang2019,Chen2015,Lin2020,Nakata2016,Efferen2021} which exhibit CDWs (left for further research when the structural relaxations are well known for these systems). A 1H monolayer TMD consists of a triangular lattice of transition metal atoms, which are connected indirectly through bonds to chalcogen atoms in a trigonal prismatic arrangement around each transition metal atom. Thus, the chalcogen atoms also form a triangular lattice above and below the triangular lattice of the transition metal atoms.

The vectors which describe the monolayer lattice are given by
\begin{equation}
\textbf{R}_1 = a_0\left(1 , 0\right) , \quad  \textbf{R}_2 = a_0\left(1/2 , -\sqrt{3}/2\right),
\end{equation}

\noindent where $a_0$ is the lattice constant of the TMD. These vectors for the monolayer are shown in Fig.~\ref{fig:Struct}(a). The transition metal atoms are denoted by filled circles, and the chalcogens are located where the lines emerging from these filled circles cross, as shown in Fig.~\ref{fig:Struct}(a). The bottom layer is always shown in red with larger filled circles for the transition metal atoms. Whilst the top layer is always shown in black, with smaller filled circles for the transition metal atoms.

In twisted 1H homobilayer TMDs, owing to the 3-fold symmetry of the monolayer, twist angles close to 0$\degree$ are distinct from those with twist angles close to 180$\degree$~\cite{Weston2020TMD,Enaldiev2020,Maity2021rec}. The former is referred to as a parallel (P) twisting and the latter anti-parallel (AP) twisting. 

In a P moir\'e~\cite{Weston2020TMD,Enaldiev2020,Maity2021rec}, as schematically shown in Fig.~\ref{fig:Struct}(e) for a marginal twist angle, a honeycomb lattice composed of two ``moir\'e sublattices'' with MX and XM stacking occurs [as shown in Fig.~\ref{fig:Struct}(c) and (d) - first letter is the bottom layer atom, with X denoting the chalcogen and M denoting the transition metal atom]. These triangular MX and XM domains comprise the majority of the area moir\'e supperlattice, since they are the lowest energy stacking. These domains are primarily connected through quasi-1D domain walls with saddle-point stacking. Neighbouring MX and XM domains are connected through a partial screw dislocation, which is equivalent to the shift of an ``in-plane bond length'' (connecting the chalcogen and transition metal atom) of the TMD in the direction of the domain wall~\cite{Enaldiev2020}. Finally, these domain walls are connected to each other through the vertex regions which have ``AA'' stacking of the bilayers. 

In contrast, a AP moir\'e~\cite{Weston2020TMD,Enaldiev2020,Maity2021rec}, as schematically shown in Fig.~\ref{fig:Struct}(f) for a marginal twist angle, has a triangular lattice composed of large hexagonal domains with 2H stacking [shown in Fig.~\ref{fig:Struct}(b)]. These hexagonal domains are connected through a screw dislocation which has a magnitude equal to the lattice constant of the TMD, in the direction of the domain wall~\cite{Enaldiev2020}. These domain walls are connected by two different vertex regions: XX and MM stacking, with the MM stacking having a slightly larger area because it is lower in energy than XX~\cite{Enaldiev2020,Naik2018}.

Note that the general conclusions drawn here are independent of the specific stacking orders of each domain. This is because the conclusions are determined from how the domains are connected and the degeneracy of the lowest energy stacking configuration. Therefore, the domains could have ``AA" stacking, for example, and similar observations will hold (the exact details of which period exhibiting which behaviour could change, however).

The moir\'e length scale~\cite{Guinea2018} is given by
\begin{equation}
    L_m = \dfrac{a_0}{2\sin(\theta/2)},
\end{equation}

\noindent where $\theta$ is the twist angle relative to 0$\degree$ and 180$\degree$ for P and AP twisting, respectively. The moir\'e lattice vectors are 
\begin{equation}
\textbf{R}^{m}_1 = L_m\left(0, 1\right) , \quad  \textbf{R}^{m}_2 = L_m\left(\sqrt{3}/2 , 1/2\right).
\end{equation}

In moir\'e materials, the issue of commensurability arises, i.e. if there is translational symmetry of the lattice on the moir\'e scale. Investigating the electronic structure usually requires commensurate structures, i.e. using DFT or atomistic tight-binding, unless a low-energy continuum model can be constructed. The requirement of commensurate moir\'e structures does not arise in our analysis. This is because the relative stacking in each domain is uniquely defined by the Burgers vector which shifts the stacking in the domain wall~\cite{Enaldiev2020}. Any slight differences in atomic positions from commensurate vs. incommensurate structures occurs through the exact areas of each type of stacking (domains, domain walls, vertex).

\subsection{Free energy}

The order parameter of the CDW phase (in a layer, $l$) is defined from the modulation in the electronic charge density~\cite{Rossnagel2011REV,Xu2021REV}, as seen by
\begin{equation}
    \rho^{(l)}(\textbf{r}) = \rho_0^{(l)}(\textbf{r}) + \delta \rho^{(l)}\left(\textbf{r},\textbf{r}_0^{(l)}\right),
\end{equation}

\noindent where the superscript $(l)$ denotes the layer, $\rho_0^{(l)}(\textbf{r})$ is the electron density in the normal state that is unit cell periodic, $\delta \rho^{(l)}(\textbf{r})$ describes the variation in the electron density from the CDW phase, with a periodicity larger than the unit cell, and $\textbf{r}_0^{(l)}$ is how the CDW is shifted relative to the lattice (more generally, when the CDWs do not coincide at a point, this shift can also depend on $j$). 

Typically, in phenomenological theories of CDWs, the CDW is expressed as
\begin{equation}
    \delta \rho^{(l)}\left(\textbf{r},\textbf{r}_0^{(l)}\right) = \sum_j \phi_j^{(l)} \textrm{Re}\Big\{e^{{i}\textbf{q}^{(l)}_j\cdot\left(\textbf{r} - \textbf{r}_{0}^{(l)}\right) }\Big\}
\end{equation}

\noindent where $\textbf{q}_j$ are the set of wavevectors which describe the CDW, and $\phi_j^{(l)}$ are the order parameters which describe the amplitude of the CDW. When there is no CDW phase $\phi_j^{(l)} = 0$, but when the phase exists this order parameter is non-zero. 

The first phenomenological theory for CDW formation in TMDs was pioneered by McMillan~\cite{McMillan1975,McMillan1976,McMillan1977}, and later studied in depth by Jacobs and Walker~\cite{Jacobs1980,Walker1981,Walker1982}. The theory is based on the Landau free energy for phase transitions, where $\phi_j^{(l)}$ is the (main) order parameter of the theory (there can also phases - how the CDWs in adjacent layers stack - which need to be found). While the free energy of each layer and all their possible interactions (those allowed by symmetry, but which do not necessarily have physical meaning) are not described in detail here, we provide a basic summary of the general form of the free energy and highlight some important terms. The readers are referred to Refs.~\citenum{McMillan1975,Jacobs1980,Walker1981,Walker1982} for a detail description of the free energy.

In addition to the usual quadratic and quartic powers of the order parameter, $\phi_j^{(l)}$, the Landau free energy also contains cubic terms which drive a weak first order phase transition (instead of a second order transition) for the onset of commensurate CDW in 2D metallic TMDs. One of these cubic terms is proportional to the product of the order parameter of each CDW component in a layer, $\phi_1^{(l)}\phi_2^{(l)}\phi_3^{(l)}$, which drives the formation of the commensurate CDW~\cite{McMillan1975}. This term persists for all periods of the CDWs with three fold symmetry. For a $3\times3$ CDW system, there are also terms which are the cube of each CDW, $[\phi_j^{(l)}]^3$~\cite{McMillan1975}. It is assumed that this later term is small and the former term dominates. Therefore, for all CDW periods, it is assumed that \textit{if one of the CDWs needs to be destroyed, all of the CDWs are destroyed, since this cubic term drives CDW formation}. In the monolayer limit, all 3 CDWs are often present. 

Here we start from a free energy which is expressed as
\begin{equation}
    \mathcal{F} = \mathcal{F}^{(1)}_{\textrm{C}} + \mathcal{F}^{(2)}_{\textrm{C}} + \mathcal{F}_{\textrm{I}}[\textbf{r}_0^{(1)},\textbf{r}_0^{(2)}]
\end{equation}

\noindent where $\mathcal{F}^{(l)}_{\textrm{C}}$ is the free energy of the CDW phase in each layer, and $\mathcal{F}_{\textrm{I}}[\textbf{r}_0^{(1)},\textbf{r}_0^{(2)}]$ is the interaction between the CDWs in each layer, as given by
\begin{equation}
\mathcal{F}_{\textrm{I}}[\textbf{r}_0^{(1)},\textbf{r}_0^{(2)}] = g\int\delta \rho^{(1)}\left(\textbf{r},\textbf{r}_0^{(1)}\right)\delta \rho^{(2)}\left(\textbf{r},\textbf{r}_0^{(2)}\right)d\textbf{r},
\label{eq:FeI}
\end{equation}

\noindent where $g$ is the coupling constant between the CDWs~\cite{Nakanishi1977}. Such a term is motivated from the electrostatic interactions between the two layers in a mean-field approximation (since a cosine charge density of one layer gives rise to a cosine electrostatic potential on the other layer~\cite{Guinea2018,Cheung2022}); a more rigorous approach would explicitly derive this interaction, and take into account screening and exchange, as done for graphene layers in Ref.~\citenum{Csaba2014}. The only terms in the interaction that do not vanish are those which share the same $\textbf{q}_j$. The relative displacements between the atomic positions modulates the energy, and causes different phase shifts between the CDWs. For a positive coupling constant (electrostatic repulsion)~\cite{Nakanishi1977}, the CDWs prefer to be out-of-phase. 

It should be noted that the exact form of the interaction is of little consequence here. All that is required for the presented analysis is that there are a finite number of possible relative stacking, connected by the Burgers vector between domains, configurations of the CDWs and there is an energy hierarchy of these different states. The exact value of the interaction strength is of little importance, provided it large. 

\begin{figure*}
\centering
\includegraphics[width=0.9\textwidth]{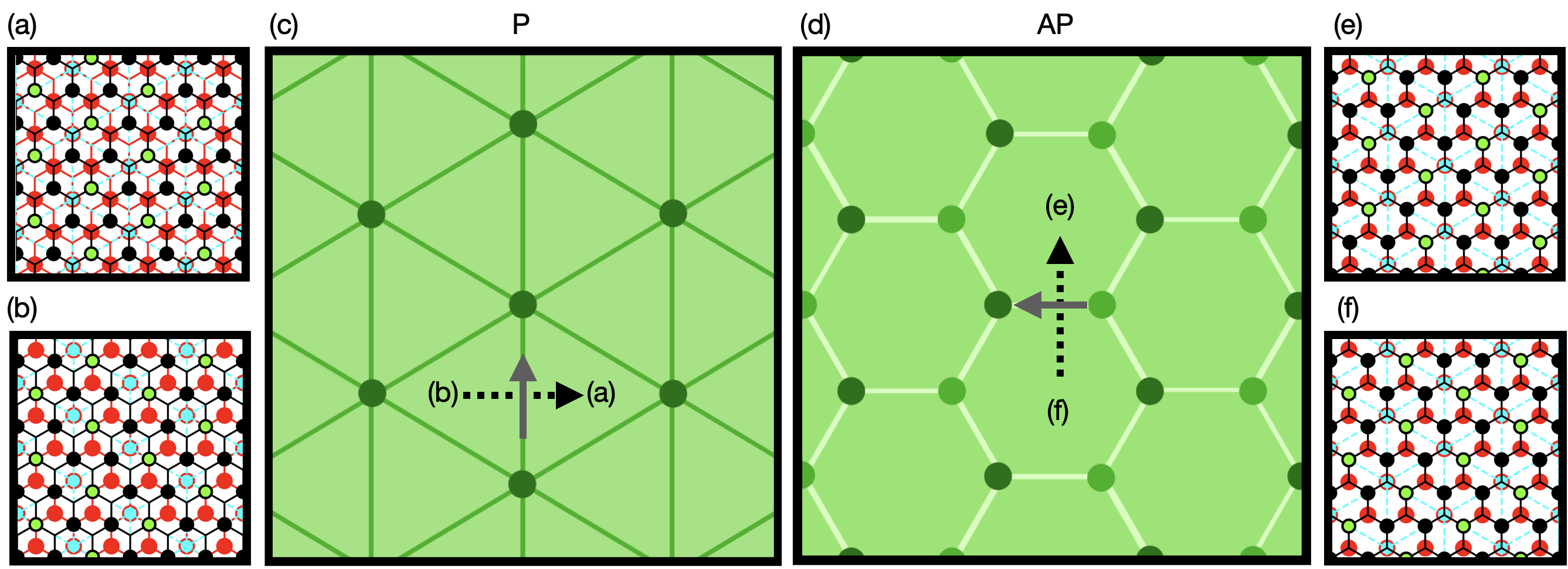}
\caption{Schematics of the $\sqrt{3}\times\sqrt{3}$ CDWs in moir\'e superlattices for P and AP twisting, as shown in the left and right panels, respectively. (a-c) - schematics of P twisting for how the $\sqrt{3}\times\sqrt{3}$ CDWs survive in moir\'e superlattices. The bottom layer is shown in red, and black is the top. Dashed blue lines denote each component of the CDW in the bottom layer. Blue circles denote where the CDW components coincide in the CDW the bottom layer, and green circles are used for the top layer. The Burgers vector in the domain wall is shown by the grey arrow, with panels (a) and (b) showing how the stacking of the bilayers and CDWs shift according to this Burgers vector.  (d-f) - AP twisting - analogous to the case of P twisting.}
\label{fig:r3br3}
\end{figure*}

\subsubsection{Effective free energy}

For marginal twist angles, as previously described, the moir\'e unit cell comprises of large domains, domain walls and vertices. The area of the vertices reaches a plateau, but the area of the domains increases as $\theta^{-2}$ (at small twist anlges), and for marginal twist angles the area of the domains dominates~\cite{Enaldiev2020}. Therefore, the vertices should not significantly contribute to the free energy, and they do not need to be accounted for in the free energy. The area of the domain walls scale as $\theta^{-1}$ for marginal twist angles, which means that they could still be significant at small twist angles, and therefore, they should be accounted for in the free energy. Moreover, domain walls only connect two neighbouring domains, which means there is a well defined Burgers vector in the domain wall. Whereas, vertex regions connect several domain walls, and therefore more domains, which means the dislocation in the vertex depends on which domains are being connected. This further motivates neglecting the vertex regions but retaining domain walls in an effective model.

Using this information, instead of solving the free energy in its full complexity, we construct an effective free energy model to describe how the CDWs survive in the moir\'e superlattice. Our minimal model for the free energy has two main contributions 
\begin{equation}
\mathcal{F}_m = \mathcal{F}_{inter} + \mathcal{F}_{intra}.
\end{equation}

\noindent The interlayer term represents the free energy of the CDWs in each layer interacting in the domains, while the intralayer term describes the free energy of the CDW existing in the domain walls. It is assumed that the CDWs exist in each domain, and that it is always the lowest energy interaction, which is a constant that is subtracted out of the free energy.

The interlayer contribution can be approximated as
\begin{equation}
    \mathcal{F}_{inter} = \mathcal{F}_{D}\sum_{ij} \cos(\textbf{q}_j\cdot\textbf{v}_{i,j})
\end{equation}

\noindent Here the summation over $i$ represents different domains, the summation over $j$ is the 3 CDWs, which are described by the vectors $\textbf{q}_j$, and $\mathcal{F}_{D}$ is free energy scale for each domain. This pre-factor is proportional to the area of the domain and the square of the magnitude of the order parameter. Finally, $\textbf{v}_{i,j}$ describes how the CDWs in each layer are shifted relative to each other
\begin{equation}
    \textbf{v}_{i,j} = \bm{\delta}_{i} + \dfrac{n_{i,j}\textbf{a}_j}{m},
\end{equation}

\noindent where $n_{i,j}$ describes the relative shift of the CDW from the freedom of placing the CDW on the supercell it creates, with $n_{i,j} = 1,..,m$ being the possible values and $m$ is the maximum possible integer, $\textbf{a}_j$ is the real-space lattice vector which describes the CDW supercell, and$\bm{\delta}_{i}$ is the intrinsic relative shift of the two layers. In the chosen convention of the theory, the Burgers vector is included in $\bm{\delta}_{i}$. For $i=0$, $\bm{\delta}_{0}$ is defined by the relative stacking of the two layers. For all other domains, the relative shift is given by
\begin{equation}
    \bm{\delta}_{i} = \bm{\delta}_{0} + \bm{b}_{0,i},
\end{equation}

\noindent where $\bm{b}_{0,i}$ is the Burgers vector which describes how the stacking of the layers changes from domain $0$ to domain $i$, which can be expressed as a sum of all the domain walls crossed $\bm{b}_{0,i} = \bm{b}_{0,1} + \bm{b}_{1,2} + ... + \bm{b}_{i-1,i}$. Note the Burgers vector is anti-symmetric $\bm{b}_{i,j} = -\bm{b}_{j,i}$.

The contribution from $\bm{\delta}_{0}$ produces a constant interlayer phase shift that is the same in each domain. The contribution from $\textbf{q}_j\cdot\bm{b}_{0i}$ produces an interlayer phase shift from the field of dislocations connecting the different domains, which is denoted by $\psi_{i,j}$, and referred to as the ``dislocation field phase shift''. This field is also fixed, and determined by the moir\'e structure and CDW period. Finally, the contribution from $n_{i,j}$ is the part which can cause different behaviour to occur. In the employed convention, for the CDW phase to survive in the domain walls, the values of $n_{i,j}$ must be the same in each domain. Therefore, the intralayer contribution to the free energy can be approximated as
\begin{equation}
    \mathcal{F}_{intra} = -\dfrac{\mathcal{F}_{W}}{2}\sum_{ii^\prime}\prod_{j} \delta_{n_{i,j}, n_{i^\prime,j}},
\end{equation}

\noindent where $\delta$ is the Kronecker delta function, and the reference system is chosen such that if the CDWs do not survive in the domain walls, the energy is 0, and if they do survive the energy is lowered by $\mathcal{F}_{W}$. In a different convention of the theory, $\bm{\delta}_{i} = \bm{\delta}_{0}$ everywhere, and the values of $n_{i,j}$ must change between domains if the CDW is to survive in the domain walls, according to $\textbf{q}_j\cdot \bm{b}_{i,i'} = \textbf{q}_j\cdot (\textbf{v}_{i,j} - \textbf{v}_{i',j}) = \textbf{q}_j\cdot \textbf{a}_j(n_{i,j} - n_{i',j})/m$.

In the employed convention, there are constraints on the allowed values of $n_{i,j}$. For example, when all of the CDW components coincide at a point (the transition metal atom), i.e. no intralayer phase shift, the first two CDW components can be placed on the lattice without constraints (provided they still cross at a transition metal atom), but the last CDW is determined by where these two CDWs cross. For example, the $\sqrt{3}\times\sqrt{3}$ CDWs to coincide at one point, it turns out that each value of $n_{i,j}$ must be the same for all $j$. 

Note that the when there is a mismatch between the CDWs in the domain walls because of the favourable stacking in the domains, it is actually only required to destroy the CDW in one of the layers. We assume that the system is at a temperature which is low enough to support the CDW phase in the domains (since interactions stabilise the phase and increase its transition temperature), but high enough that it does not persist in the monolayer. Therefore, if the CDW phase is destroyed in one layer, it is also destroyed in the other layer. The results are presented and discussed based on this assumption. The theory will also work for when the CDW is not destroyed in both layers in the domain wall, but its interpretation becomes more subtle.

\section{Results}

\begin{figure*}
\centering
\includegraphics[width=0.9\textwidth]{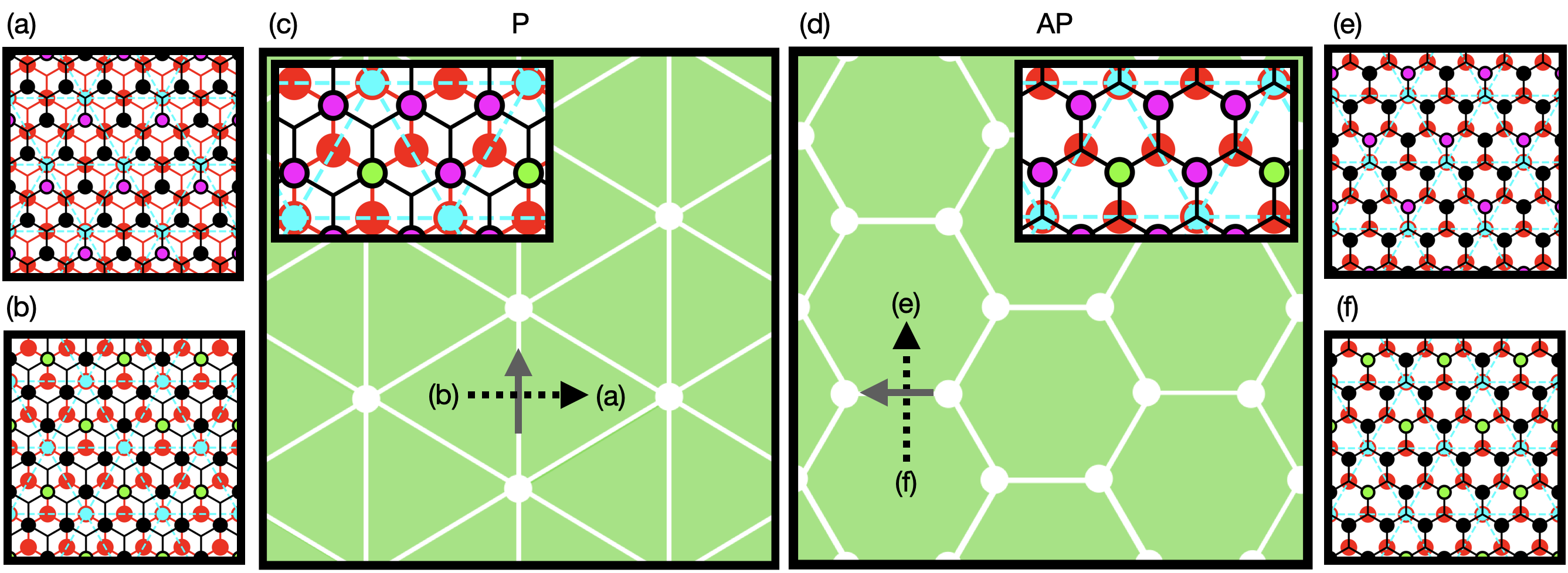}
\caption{Schematics of the $2\times2$ CDWs in moir\'e superlattices for P and AP twisting, as shown in the left and right panels, respectively. (a-c) - schematics of P twisting for how the $2\times2$ CDWs survive in moir\'e superlattices. The bottom layer is shown in red, and black is the top. Dashed blue lines denote each component of the CDW in the bottom layer. Blue circles denote where the CDW components coincide in the bottom layer, and green and pink circles are used for different stacking configurations/energies of the CDW in the top layer. The Burgers vector in the domain wall is shown by the grey arrow, with panels (a) and (b) showing how the stacking of the bilayers and CDWs shift according to this Burgers vector. The inset in panel (c) shows the 4 possible stacking arrangements of the CDWs, with the green circle denoting the lowest energy stacking and the 3 pink circles being degenerate and higher in energy. (d-f) - AP twisting - analogous to the case of P twisting.} 
\label{fig:2b2}
\end{figure*}

\subsection{$\sqrt{3}\times\sqrt{3}$ and persistence of CDWs}

The real-space lattice vectors for a $\sqrt{3}\times\sqrt{3}$ CDW are given by 
\begin{equation}
\textbf{a}_1 = a_0\left(\dfrac{3}{2} , -\dfrac{\sqrt{3}}{2}\right) , \quad  \textbf{a}_2 = a_0\left(\dfrac{3}{2} , \dfrac{\sqrt{3}}{2}\right), 
\end{equation}

\noindent where the third is $\textbf{a}_3 = \textbf{a}_2 - \textbf{a}_1$. Note these vectors can be used for both layers of P and AP twisting (since we are only concerned with where the transition metal atoms reside). Using $\textbf{a}_j\cdot\textbf{q}_{j'} = 2\pi\delta_{jj'}$, the wavevectors of the CDW can be obtained
\begin{equation}
\textbf{q}_1 = \dfrac{2\pi}{a_0}\left(\dfrac{1}{3} , -\dfrac{1}{\sqrt{3}}\right) , \quad  \textbf{q}_2 = \dfrac{2\pi}{a_0}\left(\dfrac{1}{3}, \dfrac{1}{\sqrt{3}}\right),
\end{equation}

\noindent where the third wavevector is given by $\textbf{q}_3 = \textbf{q}_1 + \textbf{q}_2$. For $\sqrt{3}\times\sqrt{3}$ CDW, $m = 3$ and this means $n_{i,j} = 1,2,3$ are the possible values. 

Taking a transition metal atom of the bottom layer (in domain $0$) to be the origin, a transition metal atom in the top layer is located at 
\begin{equation}
    \bm{\delta}_0 = a_0\left(\dfrac{1}{2}, -\dfrac{\sqrt{3}}{6}\right).
\end{equation}

\noindent for both AP and P twisting (in the XM domain - we shall come back to the MX domain later). Taking $n_{0,j} = 3$, the phase shifts of the CDWs are given by
\begin{equation}
\textbf{q}_1\cdot\bm{\delta}_{0} = \dfrac{2\pi}{3} , \quad   \textbf{q}_2\cdot\bm{\delta}_{0} = 0 , \quad  \textbf{q}_3\cdot\bm{\delta}_{0} = \dfrac{2\pi}{3}.
\end{equation}

\noindent Therefore, there is no net interaction between the CDWs of each layer, and the transition temperature of each domain should be similar to the monolayer case. 

Let us investigate how this CDW phase propagates into a neighbouring domain for AP twisting. We shall follow the path of the dashed arrow we follow the path of the dashed arrow. The initial stacking of the bilayer is described by $\bm{\delta}_0$, and is pictorially shown in Fig.~\ref{fig:r3br3}(f). Crossing the domain wall, a screw dislocation of $\textbf{b}_{0,+1} = -a_0(1,0)$ [shown by the grey arrow in Fig.~\ref{fig:r3br3}(d)] shifts the layers by a lattice constant. Therefore, the CDWs in each layer must shift by $\textbf{b}_{0,+1}$ relative to each other, as shown in Fig.~\ref{fig:r3br3}(e). The dislocation field phase shifts that are picked up are given by $\psi_{+1,1} = - 2\pi/3$, $\psi_{+1,2} = -2\pi/3$, $\psi_{+1,3} = -4\pi/3$, which also gives a net interaction of 0. Taking different values of $n_{+1,j}$ does not yield a lower energy.

This can be repeated for all possible domain walls and allowed values of $n_{i,j}$, and what is found is that the energy of each domain is identical, i.e., there are 3 degenerate stacking configurations of the CDWs, and going across domain walls inter-converts these. Therefore, the system can obtain the most energetically favourable configuration by keeping the $n_{i,j}$'s constant everywhere, which means that the CDWs are not destroyed in the domain walls, and presumably not in the vertex regions either. 

For P stacking, let us investigate what happens when going from a XM domain to a MX domain. The $\bm{\delta}_0$ describes the relative shift in the XM domain, and the relative stacking of the neighbouring MX domain is given by 
\begin{equation}
\bm{\delta}_{XM} = \bm{\delta}_{0} + \bm{b}_{0,XM} = a_0\left(1/2, \sqrt{3}/6\right).
\end{equation}

\noindent where $\textbf{b}_{0,XM} = a_0(0,1/\sqrt{3})$ shifts the lattice by an ``in-plane bond length'' [shown by the grey arrow in Fig.~\ref{fig:r3br3}(c)]. This transformation is shown in Figs.~\ref{fig:r3br3}(a) and (b). Performing a similar analysis as the AP case, it is again found that each domain has the same energy. Therefore, the CDWs in each layer are expected to persist everywhere as there is no geometrical reason why the free energy requires it to be destroyed in any domain wall. 

Figures~\ref{fig:r3br3}(c) and (d) schematically show the persistence of the $\sqrt{3}\times\sqrt{3}$ CDWs in the moir\'e supperlattice for P and AP twisting, respectively. Since the stacking of the layers is different in the domain walls and vertex regions relative to the domains, there could be a modulation in the order parameter, $\phi_j^{(l)}$, throughout the moir\'e supperlattice (different shades of green used). For example, if the saddle-point stacking in the domain wall causes a less favourable interlayer phase shift of the CDWs than the domains, the CDW won't be stabilised by the interaction as much the domains, which causes the magnitude of $\phi_j^{(l)}$ to decrease (and could even vanish) in the domain wall. The spatial variation of the order parameter, $\phi_j^{(l)}(\textbf{r})$, across a domain wall can be modelled with a sine-Gordon equation using the interaction energy of Eq.~\eqref{eq:FeI}. This is a secondary effect, however.

\subsection{$2\times2$ and destruction of CDWs}

\begin{figure*}
\centering
\includegraphics[width=0.9\textwidth]{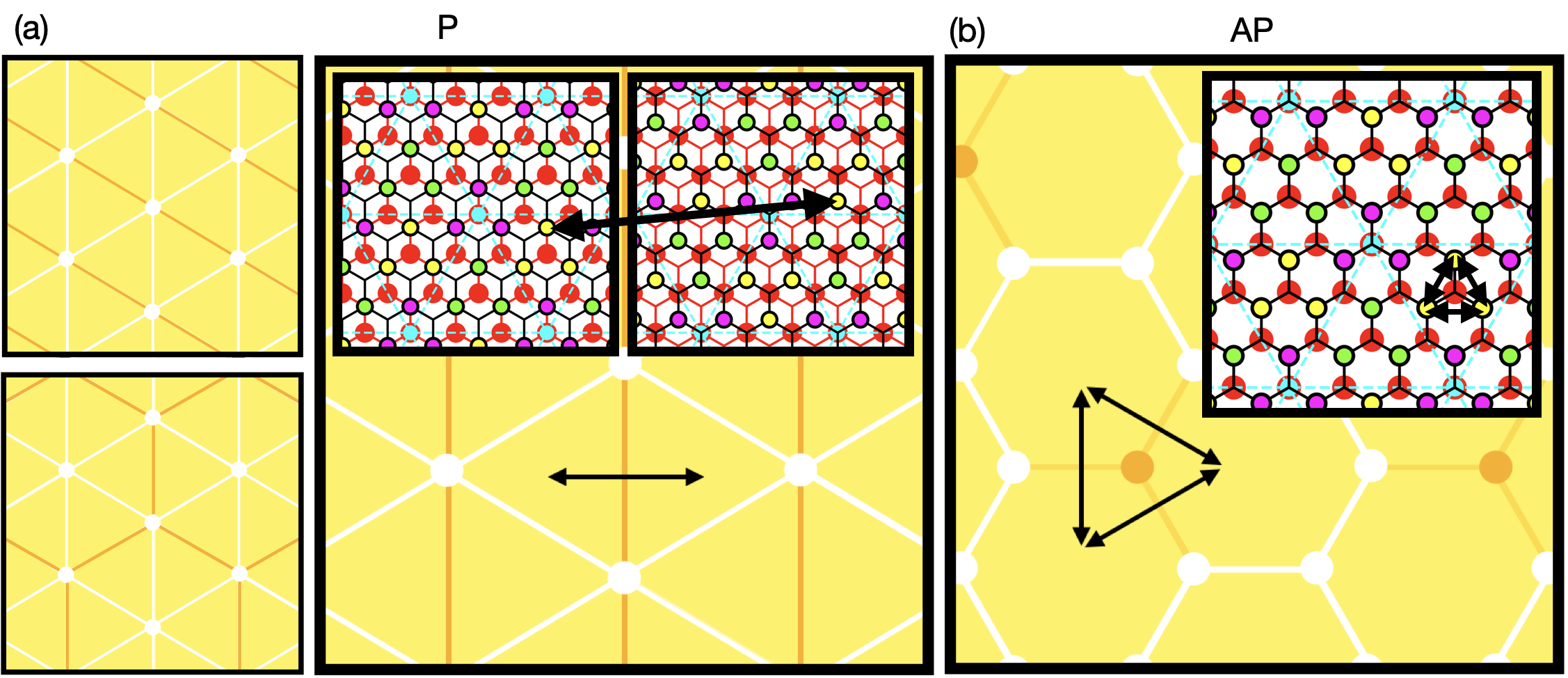}
\caption{Schematics of the $3\times3$ CDWs in moir\'e superlattices for P and AP twisting, as shown in (a) and (b), respectively. In both cases, the insets show the 9 possible stacking arrangements: the lowest state with 3 fold degeneracy is shown in yellow, the intermediate state is shown in green and is 3 fold degenerate, and finally the largest energy stacking, which is also 3 fold degenerate, is shown in pink. (a) - for P twisting, ``moir\'e dimer domains'' can form because two of the degenerate, lowest energy states are related through the Burgers vector contained in a domain wall. The arrow indicates the connected domains, where the orange domain walls are the ones which can survive. Note that several moir\'e-scale structures can emerge from these dimer-domains, as shown on the left. (b) - for AP twisting, ``moir\'e triplet domains'' can form because the 3 degenerate stacking configurations are related through the Burgers vector contained in a domain wall. These 3 domains share a vertex area and 3 domain walls, all of which could host the CDWs, since there is not geometric reason preventing it.}
\label{fig:3b3}
\end{figure*}

The real-space lattice vectors for a $2\times2$ CDW are given by 
\begin{equation}
  \textbf{a}_1 = a_0\left(1 , -\sqrt{3}\right) , \quad \textbf{a}_2 = a_0\left(1 , \sqrt{3}\right),
\end{equation}

\noindent where the third vector is given by $\textbf{a}_3 = \textbf{a}_2 + \textbf{a}_1$. The corresponding wavevectors are given by
\begin{equation}
\textbf{q}_1 = \dfrac{2\pi}{a_0}\left(\dfrac{1}{2} , -\dfrac{\sqrt{3}}{6}\right) , \quad  \textbf{q}_2 = \dfrac{2\pi}{a_0}\left(\dfrac{1}{2}, \dfrac{\sqrt{3}}{6}\right) , 
\end{equation}

\noindent where the third is given by $\textbf{q}_3 = \textbf{q}_2 - \textbf{q}_1$. In this case, $m=2$, and the allowed values of $n$ are $n_{i,j}=1,2$.

For both P and AP there are 4 possible stacking configurations of the layers in each domain, as shown in the insets of Figs.~\ref{fig:2b2}(c) and (d), respectively. One of these has an energy of $-3\mathcal{F}_D/2$ ($4\pi/6,4\pi/6,8\pi/6$), which is the lowest energy configuration, and shown in green. The other 3, which are degenerate and related by rotational symmetry, have an energy of $\mathcal{F}_D/2$ ($2\pi/6,4\pi/6,10\pi/6$), and are shown in pink. Each domain will prefer to have the most energetically favourable stacking. For neighbouring domains, there is necessarily a Burgers vector which shifts the lattice away from this most favourable interaction, as shown in Figs.~\ref{fig:2b2}(a-b) and (e-f) for P and AP twisting, respectively. Thus, provided the twist angle is small, the system must destroy the CDWs in the domain walls for there to be the most favourable stacking in each domain. Figures~\ref{fig:2b2}(c) and (d) schematically show how the order parameter of $2\times2$ CDWs is expected to exist in the moir\'e superlattice of P and AP twisting, respectively. 

\subsection{$3\times3$ and moir\'e-scale structures}

The real-space lattice vectors for a $3\times3$ CDW are given by 
\begin{equation}
  \textbf{a}_1 = a_0\left(\dfrac{3}{2} , -\dfrac{3\sqrt{3}}{2}\right) , \quad \textbf{a}_2 = a_0\left(\dfrac{3}{2} , \dfrac{3\sqrt{3}}{2}\right), 
\end{equation}

\noindent where the third vector is given by $\textbf{a}_3 = \textbf{a}_2 + \textbf{a}_1$. The corresponding wavevectors are given by
\begin{equation}
\textbf{q}_1 = \dfrac{2\pi}{a_0}\left(\dfrac{1}{3} , -\dfrac{\sqrt{3}}{9}\right) , \quad  \textbf{q}_2 = \dfrac{2\pi}{a_0}\left(\dfrac{1}{3}, \dfrac{\sqrt{3}}{9}\right), 
\end{equation}

\noindent where the third is given by $\textbf{q}_3 = \textbf{q}_2 - \textbf{q}_1$. In this case $m=3$ and $n_{i,j}=1,2,3$.

There are 9 possible stacking arrangements for $3\times3$ CDWs, but only 3 unique stacking configurations which are each 3-fold degenerate: the lowest energy configuration, $\sim-1.11\mathcal{F}_D$ ($2\pi/9,10\pi/9,10\pi/9$, for example), shown in yellow of the insets; the intermediate stacking configurations, shown in green, with energy $\sim-0.59\mathcal{F}_D$ ($4\pi/9,8\pi/9,8\pi/9$, for example); and the unfavourable configurations with energy $\sim1.70\mathcal{F}_D$ ($4\pi/9,2\pi/9,16\pi/9$). Unlike the previous cases, there is a difference between P and AP twisting.

For P twisting, starting from a domain with favourable interaction, it is possible to reach an energy equivalent state in a neighbouring domain, related by a shift of a ``bond length''. This happens when, for example, starting from the XM domain with the top yellow stacking configuration, as shown in the left inset of Fig.~\ref{fig:3b3}(a), then going to the MX domain with the bottom yellow stacking configuration (again formed by the triangle of yellow circles), as seen in the right inset of Fig.~\ref{fig:3b3}(a). Therefore, ``moir\'e dimer domains'' form when two adjacent domains have the lowest energy stacking configuration which have a relative interlayer shift that is given by the partial screw dislocation contained in the connecting domain wall. The CDW survives between these domains (remember, all 3 components of the CDW survive in this domain wall), but between different dimer domains the CDWs must be destroyed (all 3 components are destroyed). These can tile the whole moir\'e lattice, as shown in Fig.~\ref{fig:3b3}(a). Note that there are three possible ``moir\'e dimer domains'' which are related by rotational symmetry. One of them can dominate, as shown, and form nematic structures. There are two more moir\'e-scale structures of aligned dimer domains which are related by the 3-fold rotational symmetry, one of which is shown in the upper left panel of Fig.~\ref{fig:3b3}(a). Alternatively, a moir\'e-scale hexagonal structure can form from, as shown in the lower left panel of Fig.~\ref{fig:3b3}(a).

For these pristine moir\'e-scale nematic/hexagonal phases to exist, or even a single moir\'e dimer domains for that matter, it requires that neighbouring domains have the correct CDW stacking configuration for them to be connected in such a way that they can survive in the domain walls. However, if the domains initially host the CDW phases, they will presumably be randomly populated by the 3 stacking configurations. To reach one of the pristine moir\'e-scale structures shown in Fig.~\ref{fig:3b3}(a), the interlayer CDW stacking configuration in domains must be able to interchange. This might not be kinetically possible, however, as the energy barrier to convert between degenerate stacking configurations might be too large. This could result in only a few dimer domains forming, with the CDW phase being destroyed in more domain walls that necessary, since the system cannot reach the lowest energy state.

For AP twisting, the CDWs can persist through certain domain walls and form ``moir\'e -scale triplet domains'' (triangular structures) from the hexagonal domains of 2H stacking. This is schematically shown in Fig.~\ref{fig:3b3}(b), where 3 hexagonal domains which neighbour each other and share a vertex area, do not require the CDWs to be destroyed in the connecting domain walls and vertex. This occurs because the AP twisting structures have a screw dislocation of a lattice constant, and these dislocations form closed-loops, which means the CDWs can form moir\'e-scale structures with closed loops. These ``moir\'e-scale triplet domain'' structures can tile on the effective moir\'e lattice without frustration (provided the CDWs in each domain do not freeze in place in a frustrated way). Between these moir\'e-scale triangular structures, the CDW phase must be destroyed, since it converts the favourable stacking to an unfavourable one. Again, similar to the P twisting, kinetically reaching a pristine example shown in Fig.~\ref{fig:3b3}(b) might not be possible.

Using these observations, for there to be moir\'e scale ``stripe'' structures of the CDWs, there must be degenerate stacking configurations which are connected by a Burgers vector. Repeated application of this Burgers vector should cycle through degenerate states and, from translational symmetry, arrive back at the initial stacking after $m$ domains. 

If the $2\times2$ CDW has the opposite sign of the coupling constant ($g < 0$ or $\mathcal{F}_D < 0$) this stripe-phase is a possibility for AP twisting. For this case it is also possible to obtain the moir\'e-scale triplet domains. Moreover, for P twisting with $2\times2$ CDWs, moir\'e dimer domains could occur.

\subsection{$\sqrt{13}\times\sqrt{13}$ and destruction of CDWs}

Finally, the star-of-David CDWs is briefly discussed, since these are often studied in experiments~\cite{Chen2020CDW,Park2021,Ma2016,Cho2016,Tsen2015}. There are 13 possible stacking arrangements of the CDWs. Many of these are degenerate, but there is only one non-degenerate stacking with the lowest energy for both P and AP twisting. Therefore, each domain wants to be in the stacking configuration with the lowest energy, and the CDWs are necessarily destroyed in the domain walls to ensure this. This is also the case for $\sqrt{7}\times\sqrt{7}$ and $4\times4$ CDWs, as was also shown to be the case for $2\times2$ CDWs as previously discussed.

\subsection{CDWs with intralayer phase shift}

It was assumed throughout that the CDWs coincide at a single atom (in a monolayer), taken to be the transition metal atom (the same results are obtained if it is instead taken as the chalcogen atom). This is not necessarily the case, however, as Jacobs and Walker investigated~\cite{Walker1982}. The same analysis can be repeated for the above cases, but where there is an intralayer phase shift between the $j$ components of the CDW phase (note this is different to the interlayer phase different from how the CDWs in each layer stack on top of each other). While we will not go into predicting the exact details of each period again, it is expected that the examples discussed here have uncovered the possible cases for how CDWs are modulated by moir\'e superlattices (as summarised in Fig.~\ref{fig:summ} and Tab.~\ref{tab:summ}, surviving everywhere, destroyed in domains, and moir\'e-scale structures).

\section{Discussion}

%
%

It has been found in marginally twisted double bilayer graphene (two Bernal stacked bilayer graphene twisted relative to each other) that the electronic structure of the large domains corresponds well to the ABAB and ABCA stacked graphene multilayer~\cite{Kerelsky2020ABCA}. Therefore, a marginal twist angle twisted bilayer of metallic TMDs at small twist angles would, most likely, inherit the properties of the bilayer with the stacking sequence of the domain. If this bilayer has distinct CDW phases from the monolayer, not just being modulated from the interaction of the two layers, then the domains of the marginal twisted metallic TMDs should resemble these bilayer CDW phases. Moving away from the domains, across domain walls or the vertex regions, would presumably destroyed these phases. Therefore, a similar physical picture to $2\times2$ CDWs could occur if the electronic structure/electron phonon coupling of the bilayer is significantly different to the monolayer.

An assumption of the presented analysis is that the moir\'e scale atomic reconstruction does not change between the normal state and the CDW phases. As CDWs can couple strongly to the lattice, this assumption might not always hold, and so we can speculate what would happen. If the largest domains do not have the lowest energy stacking, but actually the vertex areas with AA/XX/MM stacking have the lowest energy in the CDW phase, it could drive the area of these regions to increase. Therefore, the moir\'e reconstruction could depend on which phase the TMD is in. This is conceptually similar to the changes in the domain sizes of the P twisted semiconducting TMDs in an electric field, driven by their ferroelectric properties~\cite{Weston2022,Enaldiev2022Ferro}. Development of a continuum elasticity field theory, based on that of Ref.~\citenum{Enaldiev2020}, which accounts for the different energies of the CDW phase and normal state in different stacking configurations would be required.


It has also been shown that CDW phases are sensitive to strain~\cite{Gao2018,Zhang201743a,Soumyanarayanan20131D}. Strain has been shown to alter the CDW to a non-3-fold symmetric CDW~\cite{Gao2018,Zhang201743a,Soumyanarayanan20131D}, or to increase/decrease the transition temperature~\cite{Xu2021REV}. In small twist angle moir\'e structures, the strain is primarily located at near the domain walls~\cite{Wijk2015}. Therefore, it could be expected that some modification of the CDW occurs in these regions where all of the strain is located. 


\section{Outlook}

The results presented here (see Preview Section for summary) are offered as a prediction of how the CDW phases of different periods survive in moir\'e supperlattices of marginal twist angle metallic TMDs. These predictions can be verified by performed scanning tunnelling experiments to observe how the CDWs phases are modulated or destroyed over the domain walls of moir\'e supperlattices. In scanning tunnelling spectroscopic techniques, the doping level can be altered using a back gate. Here we had to assume the electronic structure of the monolayers is not significantly altered, and therefore, the predictions made here could be applicable to doping levels away from where flat bands form. For doping levels in the flat bands, if they form in metallic TMDs, \textit{ab initio} calculations are required to make accurate predictions.

While we have focused on twisted homobilayers, the presented theory could also work for monolayers twisted on multilayers through to bulk crystals, e.g. mono-bilayers, mono-trilayers, ... etc., provided the layer in contact with the twisted monolayer has the same CDW phase (period). Moreover, the presented theory could also be applied to heterobilayers (or the previously mentioned moir\'e structures), provided the lattice constant and CDW period is similar enough to create large domains. We hope the theory presented here will motivate further study of how monolayers susceptible to CDW formation are modulated by moir\'e supperlattices, or if emergent behaviour dominates. While we have focused on 1H monolayers of TMDs, the 1T polytype would naturally be the next system to investigate, and then other monolayers which are susceptible to CDWs. Finally, developing continuum models which couple the periodic lattice distortion and relative areas of each domain in the CDW/normal phase could yield CDW-driven structural transitions in TMDs, similar to the ferroelectric behaviour of semiconducting TMDs with parallel twisting. 

\section{Acknowledgements}

We thank Vladimir Enaldiev and James Mchugh for stimulating discussions. This work was supported by EC-FET European Graphene Flagship Core3 Project, EPSRC grants EP/S030719/1 and EP/V007033/1, and the Lloyd Register Foundation Nanotechnology Grant.

\bibliographystyle{apsrev4-1}
\bibliography{REF}

\end{document}